\documentclass[%
 reprint,
superscriptaddress,
nofootinbib,
 amsmath,amssymb,
 aps,
]{revtex4-1}
\usepackage{graphics,epsfig,psfrag}
\usepackage{color}
\usepackage{graphicx}
\usepackage{dcolumn}
\usepackage{bm}
\usepackage{hyperref}

\begin{document}

\title{Deflection of light rays\\
by spherically symmetric black hole in dispersive medium}

\author{Oleg Yu. Tsupko}
\email{tsupko@iki.rssi.ru; tsupkooleg@gmail.com}
\affiliation{Space Research Institute of Russian Academy of Sciences, \\
Profsoyuznaya 84/32, Moscow 117997, Russia}
\affiliation{Institute for Advanced Study (Hanse-Wissenschaftskolleg), Lehmkuhlenbusch 4, 27753 Delmenhorst, Germany}

\date{\today}

\begin{abstract}
The influence of the medium on the gravitational deflection of light rays is widely discussed in literature for the simplest non-trivial case: cold non-magnetized plasma. In this article, we generalize these studies to the case of an arbitrary transparent dispersive medium with a given refractive index. We calculate the deflection angle of light ray moving in a general spherically symmetric metric in the presence of medium with the spherically symmetric refractive index. The equation for the radius of circular light orbits is also derived. We discuss in detail the properties of these results and various special cases. In particular, we show that multiplying the refractive index by a constant does not affect the deflection angle and radius of circular orbits. At the same time, the presence of dispersion makes the trajectories different from the case of vacuum even in spatially homogeneous medium. As one of the applications of our results, we calculate the correction to the angle of vacuum gravitational deflection for the case when a massive object is surrounded by homogeneous but dispersive medium. As another application, we present the calculation of the shadow of a black hole surrounded by medium with arbitrary refractive index. Our results can serve as a basis for studies of various plasma models beyond the cold plasma case.
\end{abstract}

\maketitle

\section{Introduction}

Light rays moving in an inhomogeneous medium propagate along curved paths. Similarly, if a gravitating body is surrounded by some kind of medium, the trajectories of the rays will differ from the case of motion near the gravitating body in a vacuum. In the presence of medium around the gravitating body, the deflection of light rays will be determined by a complex combination of various effects: gravity, refraction, dispersion.

The theory of the propagation of light rays in curved spacetime in the presence of a medium was developed in the classic monograph by Synge \cite{Synge-1960}, see also 
papers \cite{Bicak-1975, Breuer-Ehlers-1980, Kulsrud-Loeb-1992}. The light deflection in the linearized regime, when the total deflection can be calculated as the sum of the vacuum gravitational deflection and refraction in an inhomogeneous medium, was discussed for the case of plasma, e.g., in works \cite{Muhleman-Johnston-1966, Bliokh-Minakov-1989}. In his monograph on ray optics in General Relativity, Perlick \cite{Perlick-2000} derived an integral formula for the exact deflection angle of a light ray moving in the equatorial plane of Kerr black hole surrounded by a spherically symmetric plasma distribution.

Recent studies of the influence of medium on the propagation of rays in a gravitational field are mainly devoted to the various effects of gravitational lensing in the presence of plasma, which fills cosmic space, and is also concentrated around compact objects. In the works of Bisnovatyi-Kogan and Tsupko \cite{BK-Tsupko-2009, BK-Tsupko-2010}, we investigated the case of weak deflection of photons in the Schwarzschild metric in the presence of plasma. These results were further generalized to the case of weak deflection in the Kerr metric \cite{Morozova-2013}. In the work of Tsupko and Bisnovatyi-Kogan \cite{Tsupko-BK-2013}, gravitational lensing in plasma in case of a strong deflection was investigated. In particular, the deflection angles were calculated in the strong deflection limit, and the properties of high-order images \cite{Virbhadra-2000, Bozza-2001} were calculated for the case of a homogeneous plasma. Er and Mao \cite{Er-Mao-2014} numerically examined strong lens systems in the presence of plasma. Various astrophysical situations associated with strong deflection of light rays near compact objects in the presence of plasma were studied in a series of works by Rogers \cite{Rogers-2015, Rogers-2017a, Rogers-2017b}. Perlick, Tsupko and Bisnovatyi-Kogan \cite{Perlick-Tsupko-BK-2015} have fully analytically investigated the shadow of spherically symmetric black holes in the presence of plasma. In the next article, we analytically investigated the effect of plasma on light ray propagation and on formation of the shadow for Kerr black hole \cite{Perlick-Tsupko-2017}. Later these issues were further discussed in articles of Yan \cite{Yan-2018}, Huang, Dong and Liu \cite{Huang-2018}, Kimpson, Wu and Zane \cite{Kimpson-2019a, Kimpson-2019b}. Different problems of light deflection in presence of plasma have been recently studied in series of papers of Crisnejo and Gallo with coauthors \cite{Crisnejo-Gallo-2018, Crisnejo-Gallo-Rogers-2019, Crisnejo-Gallo-Villanueva-2019, Crisnejo-Gallo-Jusufi-2019}. In particular, Crisnejo, Gallo and Jusufi \cite{Crisnejo-Gallo-Jusufi-2019} have calculated the higher order terms of the deflection angles.
Wave effects in presence of Solar gravity and Solar corona were studied in works of Turyshev and Toth \cite{Turyshev-2019a, Turyshev-2019b}. In a recent paper \cite{Tsupko-BK-2020}, we have investigated gravitational microlensing in the presence of plasma (`hill-hole' effect). For some other recent studies see \cite{Perlick-2017, Er-Rogers-2018, Er-Rogers-2019, Sareny-2019, Jin-2020, Wagner-Er-2020, Fathi-2020, Matsuno-2020}. Recent review of plasma effects in gravitational lensing can be found in \cite{BK-Tsupko-Universe-2017}.

In most of the mentioned works, the simplest plasma model is considered -- cold non-magnetized plasma. In this case, the plasma has a simple expression for refractive index, and many formulas take on a relatively simple form. In this paper, we go beyond this approximation and consider a medium with an arbitrary refractive index. The formulas we obtained make it possible to easily calculate the deflection angles, circular orbits and the shadow of a black hole for any known spherically symmetric refractive index. For example, these can be more complex plasma models.

The paper is organized as follows. In Sec.II, we briefly describe the relativistic geometrical optics in medium on basis of Synge's approach. In Sec.III, we calculate the deflection angle for light ray moving in spherically symmetric spacetime in presence of spherically symmetric medium. In Sec. IV, we find the equation for the circular light orbits. In Sec.\ref{sec:discussion}, we present the discussion of our results and consider various particular cases. As an application of our results, in Sec.\ref{sec:correction-disp} we calculate the correction to vacuum gravitational deflection due to presence of homogeneous dispersive medium. As another application, in Sec.\ref{sec:shadow} the angular radius of shadow of black hole surrounded by transparent spherically symmetric medium is analytically calculated. Concluding remarks are presented in Sec.\ref{sec:consclusions}.

We use units such that $G = c= 1$, and signature is $\{-,+,+,+\}$. Latin indices take the values $i, k = 0,1,2,3$, whereas Greek indices take the values $\alpha, \beta = 1,2,3$. We denote the differentiation with respect to affine parameter by dot and the differentiation with respect to radial coordinate by prime.

\section{General relativistic geometrical optics in dispersive medium}

General relativistic geometrical optics in curved spacetime filled with isotropic transparent medium was developed by Synge \cite{Synge-1960}. Let us consider spacetime with metric coefficients $g_{ik}$ which are known functions of coordinates. In this spacetime, we consider medium specified by its refractive index $n$ (which is reciprocal of the phase-speed) and its four-velocity $V^i$. Refractive index $n$ is considered as a given function of the coordinates $x^i$ and photon frequency $\omega$; four-velocity $V^i$ of medium is a given function of coordinates. Phase-speed and frequency are measured in the instantaneous rest frame of the medium.

In Synge's approach, relativistic geometric optics is based on so-called medium equation \cite{Synge-1960}:
\begin{equation} \label{medium-equation}
n^2 = 1 + \frac{p_i p^i}{\left(p_k V^k \right)^2} \, .
\end{equation}
The photon frequency $\omega(x^i)$ measured by observer at position $x^i$ is related with $p_i$ and $V^i$ by the following formula:
\begin{equation} \label{synge-freq-general}
p_i V^i  = - \omega(x^i) \, .
\end{equation}

To apply the Hamiltonian method, the medium-equation is rewritten in the form,
\begin{equation}
H(x^i,p_i) = 0 \, ,   
\end{equation}
where the Hamiltonian is
\begin{equation} \label{Hamiltonian-1}
H(x^i,p_i) = \frac{1}{2} \left\{ g^{ik} p_i p_k - (n^2-1) \left(p_k V^k \right)^2 \right\}  \, .
\end{equation}
Then, propagation of the light rays is described by Hamilton’s equations:
\begin{equation} \label{Ham-equations}
\dot{x}^i = \frac{\partial H}{\partial p_i}  \, , \; \; \dot{p}_i  = - \frac{\partial H}{\partial x^i} \, ,
\end{equation}
where a dot means differentiation with respect to an affine parameter $\lambda$ changing along the light trajectory.

Let us now consider static spacetime,
\begin{equation}
g_{ik} dx^i dx^k = g_{\alpha \beta} dx^\alpha dx^\beta + g_{00} (dx^0  )^2 ,  
\end{equation}
where the coefficients $g_{ik}$ are independent of $t$. Let us also consider the static medium, so medium refractive index $n$ is function of frequency $\omega$ and space-coordinates $x^\alpha$ but is independent of $t$. Therefore, the whole Hamiltonian is independent of $t$. In such a case, from equation for $p_0$ from (\ref{Ham-equations}) we read that $p_0 = \mbox{const}$.

Additionally, for static medium we have \cite{Synge-1960}
\begin{equation}
V^\alpha = 0 \, , \quad V^0 = \sqrt{-g^{00}} \, ,   
\end{equation}
and therefore the relation (\ref{synge-freq-general}) is simplified to
\begin{equation} \label{synge-freq-static}
\omega(x^\alpha) = - p_0 \sqrt{-g^{00}} \, .
\end{equation}

Using (\ref{synge-freq-static}), we write the Hamiltonian (\ref{Hamiltonian-1}) as
\begin{equation} \label{Hamiltonian-n}
H(x^i,p_i) =  \frac{1}{2} \left\{ g^{ik} p_i p_k - (n^2-1)
\left(p_0 \sqrt{-g^{00}}\right)^2 \right\} =
\end{equation}
\[
= \frac{1}{2} \left\{  g^{00} p_0 p_0 + g^{\alpha \beta} p_\alpha
p_\beta + (n^2-1) p_0^2 \, g^{00} \right\} =
\]
\[
= \frac{1}{2} \left\{ 
g^{\alpha \beta} p_\alpha p_\beta + n^2 g^{00} p_0^2 \right\} .
\]

\section{Deflection angle in spherically symmetric static spacetime filled by spherically symmetric medium}

In this Section we derive the exact integral expression for the deflection angle of light ray moving in general spherically symmetric static spacetime filled by transparent static medium with spherically symmetric refractive index $n$.

We consider spherically symmetric and static spacetime:
\begin{equation}
g_{ik} dx^i dx^k = - A(r) \, dt^2 + B(r) \, dr^2 
\label{metric}
\end{equation}
\[
+ \, D(r) \left( d \vartheta^2 + \sin^2 \vartheta \, d \varphi^2 \right) \, ,
\]
where coefficients $A(r)$, $B(r)$ and $D(r)$ are positive and independent of $t$. In this spacetime we consider spherically symmetric static medium which refractive index is a given function of radial coordinate $r$ and photon frequency $\omega$: $n(\omega,r)$.

Due to spherical symmetry, we restrict ourselves by motion in equatorial plane: $\vartheta = \pi/2$, $p_\vartheta=0$. Then, Hamiltonian (\ref{Hamiltonian-n}) is simplified to:
\begin{equation}
H(x^i, p_i) = \frac{1}{2} \left\{ \frac{p_r^2}{B(r)} + \frac{p_\varphi^2}{D(r)} - \frac{n^2 p_0^2}{A(r)}  \right\}  .
\end{equation}

For the frequency $\omega(r)$ measured by static observer at position $r$ we obtain from eq.(\ref{synge-freq-static}):
\begin{equation}
\omega(r) = - \frac{p_0}{\sqrt{A(r)}} \, ,    
\end{equation}
where $p_0$ is the constant of motion. If we further consider only asymptotically flat spacetimes, then $A(r) \to 1$ for $r \to \infty$. Then we have:
\begin{equation}
p_0 = - \omega_0 \, ,    
\end{equation}
where $\omega_0 \equiv \omega(\infty)$. Therefore for the frequency $\omega(r)$ we have gravitational redshift formula
\begin{equation} \label{freq-relation}
\omega(r) = \frac{\omega_0}{\sqrt{A(r)}}    \, .
\end{equation}

Equations of motion:
\begin{equation} \label{p-varphi}
\dot{p}_\varphi = - \frac{\partial H}{\partial \varphi} = 0 \, ,
\end{equation}
\begin{equation} \label{eq-pr-dot}
\dot{p}_r = - \frac{\partial H}{\partial r} = 
\frac{1}{2} \left\{ 
\frac{p_r^2 B'(r)}{B^2(r)} + \frac{p_\varphi^2 D'(r)}{D^2(r)} + \right.
\end{equation}
\[
\left. + \frac{(n^2)'\omega_0^2}{A(r)} - \frac{n^2 \omega_0^2 A'(r)}{A^2(r)} 
\right\}
\, ,
\]
\begin{equation}
\dot{\varphi} = \frac{\partial H}{\partial p_\varphi} = \frac{p_\varphi}{D(r)}    \, ,
\end{equation}
\begin{equation} \label{eq-r-dot}
\dot{r} = \frac{\partial H}{\partial p_r} = \frac{p_r}{B(r)} \, ,
\end{equation}
with $H=0$, i.e.
\begin{equation} \label{H-eq-0}
\frac{p_r^2}{B(r)} + \frac{p_\varphi^2}{D(r)} - \frac{n^2 \omega_0^2}{A(r)} = 0
\end{equation}
From eq.(\ref{p-varphi}) we find that $p_\varphi$ is the constant of motion.

To derive the orbit equation, we write
\begin{equation}
\frac{d\varphi}{dr} = \frac{\dot{\varphi}}{\dot{r}} = \frac{p_\varphi \, B(r)}{p_r \, D(r)}   \, .  
\end{equation}

Expressing $p_r$ from (\ref{H-eq-0}), we find:
\begin{equation} \label{eq-orbit-general}
\frac{d \varphi}{dr} = \pm \frac{\sqrt{B(r)}}{\sqrt{D(r)}} \left( \frac{\omega_0^2}{p_\varphi^2} h^2(r) - 1 \right)^{-1/2}    \, .
\end{equation}
Here we have introduced the function
\begin{equation} \label{h-def}
h^2(r) = \frac{D(r)}{A(r)} \, n^2(\omega(r),r) \, .    
\end{equation}
Function (\ref{h-def}) is generalization of function $h(r)$ used in \cite{Perlick-2000, Tsupko-BK-2013, Perlick-Tsupko-BK-2015} for cold plasma. In particular case of the photon motion near Schwarzschild black hole in vacuum, function (\ref{h-def}) coincides with `effective potential of photon' introduced in book of Misner, Thorne, Wheeler \cite{MTW-1973}.

\begin{figure}
\begin{center}
\includegraphics[width=0.48\textwidth]{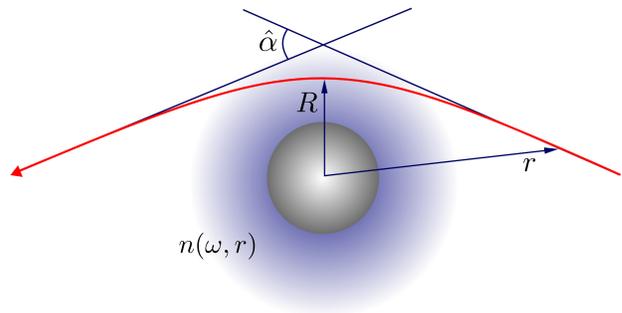}
\end{center}
\caption{Deflection angle $\hat{\alpha}$ of light ray moving near the black hole surrounded by transparent dispersive medium with the refractive index $n(\omega,r)$. The distance of the closest approach $R$ is the minimum value of the radial coordinate $r$ for this trajectory.}
\label{fig:deflection}
\end{figure}

Let us consider the situation when the trajectory of a photon passes through the point of closest approach. Then it is convenient to express everything in terms of the distance of the closest approach $R$ (minimal value of the coordinate $r$). As $R$ corresponds to the turning point of the trajectory, we should have $dr/d\varphi = 0$ if $r=R$. From (\ref{eq-orbit-general}) we find:
\begin{equation}
h^2(R) = \frac{p_\varphi^2}{\omega_0^2} \, .
\end{equation}
Then the orbit equation (\ref{eq-orbit-general}) takes form:
\begin{equation}
\label{eq-orbit-R}
\frac{d \varphi}{dr} = \pm \frac{\sqrt{B(r)}}{\sqrt{D(r)}} \left( \frac{h^2(r)}{h^2(R)} - 1 \right)^{-1/2}    \, .
\end{equation}

Now we calculate the deflection angle $\hat{\alpha}$ of light ray moving from infinity towards the center, reaching the distance of the closest approach $R$ and again flying away to infinity (Fig.\ref{fig:deflection}). Without loss of generality, we will assume that the ray moves in such a way that its angular coordinate $\varphi$ increases. Then, when moving towards the center, the coordinate $r$ is decreasing, and one need to use `minus' sign in the equation (\ref{eq-orbit-R}), and when moving from the center to infinity, `plus' sign should be used. Change the angular coordinate can be written as follows:
\begin{equation}
\Delta \varphi = 2 \int \limits_R^\infty 
\frac{\sqrt{B(r)}}{\sqrt{D(r)}} \left( \frac{h^2(r)}{h^2(R)} - 1 \right)^{-1/2}   dr \, .
\end{equation}
Since motion along a straight line corresponds to a change $\Delta \varphi = \pi$, then $\pi$ must be subtracted to obtain the deflection angle $\hat{\alpha}$:
\begin{equation} \label{alpha}
\hat{\alpha} = 2 \int \limits_R^\infty 
\frac{\sqrt{B(r)}}{\sqrt{D(r)}} \left( \frac{h^2(r)}{h^2(R)} - 1 \right)^{-1/2} dr \, - \, \pi  \, .
\end{equation}
Formula (\ref{alpha}) with (\ref{h-def}) allows to calculate the deflection angle in any spherically symmetric spacetime in presence of spherically symmetric medium defined by known refractive index. This formula is generalization of plasma formula in \cite{Perlick-Tsupko-BK-2015}. We have managed to write the formula for a more general case in the same form as in \cite{Perlick-Tsupko-BK-2015}, reducing all the differences to a more general definition of a function $h(r)$ only.

To apply the formula (\ref{alpha}) for some medium with given refractive index $n(\omega,r)$, one should use $\omega$ in the form of $\omega(r)$ from (\ref{freq-relation}). Then $n(\omega(r),r)$ should be substituted to eq.(\ref{h-def}) to obtain $h(r)$. The function $h(r)$ will also contain the frequency at infinity $\omega_0$. To get $h(R)$, one should take the refractive index as $n(\omega(R),R)$.

For example, cold non-magnetized plasma has the index of refraction:
\begin{equation}\label{refr-index-plasma-0}
n^2 = 1 - \frac{\omega_p^2(r)}{\omega^2} \, ,   
\end{equation}
where $\omega_p$ is electron plasma frequency and $\omega$ is photon frequency.
In presence of gravity we have:
\begin{equation}\label{refr-index-plasma}
n^2 = 1 - \frac{\omega_p^2(r)}{\omega^2(r)} = 1 - A(r) \frac{\omega_p^2(r)}{\omega_0^2} \, .    
\end{equation}
Therefore
\begin{equation}
h^2(r) = \frac{D(r)}{A(r)} \left( 1 - A(r) \frac{\omega_p^2(r)}{\omega_0^2} \right) \, ,
\end{equation}
and
\begin{equation}
h^2(R) = \frac{D(R)}{A(R)} \left( 1 - A(R) \frac{\omega_p^2(R)}{\omega_0^2} \right) \, .
\end{equation}

\section{Circular light orbits}

Along a circular light orbit, it should be $\dot{r}=0$ and $\ddot{r}=0$. Due to relation (\ref{eq-r-dot}), $\dot{r}=0$ leads to $p_r=0$. From (\ref{H-eq-0}) we obtain:
\begin{equation} \label{circ-01}
\frac{p_\varphi^2}{D(r)} = \frac{n^2 \omega_0^2}{A(r)} \, .
\end{equation}

Additionally, from (\ref{eq-r-dot}), we read:
\begin{equation} 
\dot{p}_r =  \frac{d}{d \lambda} (B(r) \dot{r})   = B'(r) \, \dot{r}^2 + B(r) \, \ddot{r} \, .
\end{equation}

Therefore, two conditions, $\dot{r}=0$ and $\ddot{r}=0$, together lead to $\dot{p}_r = 0$. Using eq.(\ref{eq-pr-dot}), we find the second equation for circular light orbits:
\begin{equation} \label{circ-02}
\frac{p_\varphi^2 D'(r)}{D^2(r)} + \frac{(n^2)'\omega_0^2}{A(r)} - \frac{n^2 \omega_0^2 A'(r)}{A^2(r)} = 0    \, .
\end{equation}
Substituting $p_\varphi^2$ from (\ref{circ-01}) into (\ref{circ-02}), we find that the equation for the radius of a circular light orbit can be compactified to the following simple form:
\begin{equation} \label{eq-circ}
\frac{d}{dr} h^2(r) = 0 \, .
\end{equation}
Analogous equation for cold plasma case was derived in our previous paper \cite{Perlick-Tsupko-BK-2015}.

\section{Discussion and particular cases}
\label{sec:discussion}

Two properties of the results obtained can be distinguished:

(i) First, since the refractive index is represented in the formulas  as a ratio, it is clear that multiplying the refractive index by a constant (i.e., replacing $n$ by $C_0 n$) does not change the deflection angle. Note that the speed of ray propagation in this case, of course, will change by a factor of $C_0 $. In the same way, it is seen that such a change will not affect the solutions of the equation (\ref{eq-circ}), i.e. on the radii of circular orbits. In the absence of gravity, this property is well known: the equations of ray trajectories describing the refraction in an optically inhomogeneous medium include quantities of the type $\nabla n/n$.

(ii) At the same time, in the presence of a gravitational field, a new property appears. If the medium is dispersive, but at the same time spatially homogeneous, then in the absence of a gravitational field the ray will not bend. The situation changes with the addition of a gravitational field. Since the photon frequency in the gravitational field changes according (\ref{freq-relation}), then, even in the case of a spatially homogeneous medium, a dependence on spatial coordinates, i.e. effective non-homogeneity, appears: $n(\omega(r))$. As a result, the ray trajectories in the dispersive medium will differ from the vacuum ones, even if the medium is homogeneous. This property was already mentioned in our previous papers with particular attention to homogeneous plasma case. Here we find the deflection angle (\ref{alpha}) for general case of dispersive medium, and below we will also calculate the correction to vacuum case.\\

Let us further consider the particular cases:

\textit{Vacuum or homogeneous medium in absence of gravity.}

\begin{figure}
\begin{center}
\includegraphics[width=0.48\textwidth]{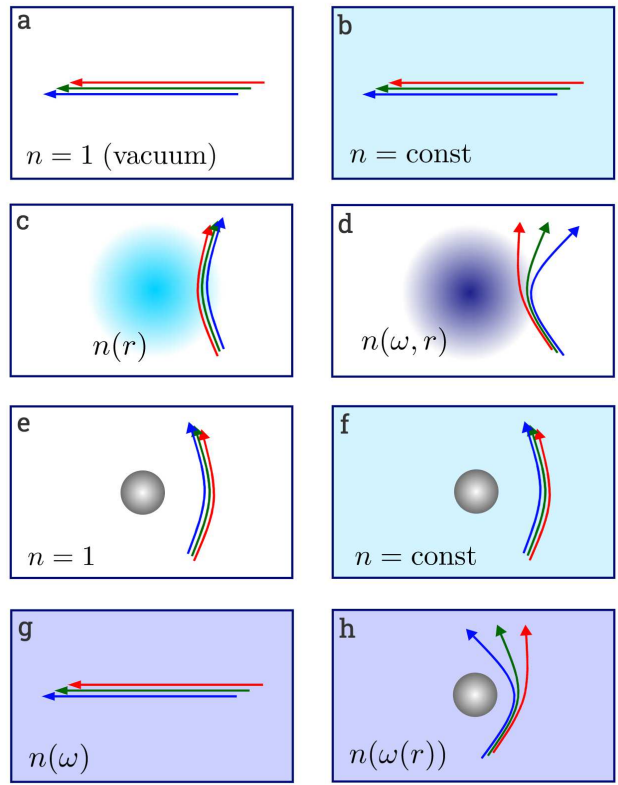}
\end{center}
\caption{Comparison of light deflection in different media, see discussion in Sec.\ref{sec:discussion}. a) Light rays of three different frequencies in vacuum, without gravity. Rays are not deflected. b) Homogeneous non-dispersive medium in absence of gravity. There is no deflection. c) Non-homogeneous non-dispersive medium in absence of gravity. There is achromatic deflection (refraction), because the refractive index depends on space coordinate. d) Non-homogeneous dispersive medium in absence of gravity. There is chromatic refraction, because the refractive index depends on both $\omega$ and $r$. e) Achromatic gravitational deflection in vacuum. f) Achromatic deflection near a gravitating body surrounded by homogeneous non-dispersive medium. Trajectories are the same as in previous case, although light speed is smaller. g) Homogeneous dispersive medium in absence of gravity. Rays of any frequency are not deflected due to the uniformity of the medium (no refraction). h) Homogeneous dispersive medium in presence of gravity. In comparison with previous case, the deflection occurs, because the photon frequency in gravitational field changes according (\ref{freq-relation}), and the medium becomes effectively non-homogeneous. This deflection is chromatic.}
\label{fig:table}
\end{figure}

This is trivial case: the deflection is absent in vacuum (Fig.\ref{fig:table}a) and in homogeneous medium, both in non-dispersive (Fig.\ref{fig:table}b) or dispersive (Fig.\ref{fig:table}g). Taking $A(r)=B(r)=1$, $D(r)=r^2$ and $n=\mbox{const}$ in (\ref{alpha}), we find:
\begin{equation} 
\hat{\alpha} = 2 \int \limits_R^\infty 
\left( \frac{r^2}{R^2} - 1 \right)^{-1/2} \frac{dr}{r} \, - \, \pi  \, = \pi - \pi = 0 \, .
\end{equation}
The use $n(\omega)$ does not change the result.\\

\textit{Non-homogeneous non-dispersive medium, in absence of gravity}, see Fig.\ref{fig:table}c.

Taking $A(r)=B(r)=1$, $D(r)=r^2$ and $n=n(r)$ in (\ref{alpha}), we find:
\begin{equation} \label{alpha-non-gravity}
\hat{\alpha} = 2 \int \limits_R^\infty 
\left( \frac{r^2 n^2(r)}{R^2 n^2(R)} - 1 \right)^{-1/2} \frac{dr}{r} \, - \, \pi  \, .
\end{equation}
For propagation in absence of gravity, this formula is known, compare, for example, with Born and Wolf \cite{Born-Wolf-1959}, Zheleznyakov \cite{Zheleznyakov-1996}.\\

\textit{Non-homogeneous dispersive medium, in absence of gravity}, see Fig.\ref{fig:table}d.

It means that refractive index depends on the space coordinates and the photon frequency: $n = n(\omega,r)$. In this case we have refractive deflection, which is calculated with formulae (\ref{alpha-non-gravity}) with the only difference that the refractive index now depends on $\omega$ also, and the trajectory and deflection angle will be different for different $\omega$.\\

\textit{Gravitational deflection in vacuum}, see Fig.\ref{fig:table}e.

Taking $n=1$ in (\ref{alpha}), we obtain the deflection angle of light ray moving in spherically symmetric metric (\ref{metric}), compare with (8.5.6) of Weinberg \cite{Weinberg-1972}:
\begin{equation} \label{alpha-vacuum}
\hat{\alpha} = 2 \int \limits_R^\infty 
\frac{\sqrt{B(r)}}{\sqrt{D(r)}} \left( \frac{D(r)A(R)}{A(r)D(R)} - 1 \right)^{-1/2} dr \, - \, \pi  \, .
\end{equation}

\textit{Homogeneous non-dispersive medium, in presence of gravity}, see Fig.\ref{fig:table}f.

It means that refractive index does not depend on the photon frequency and space coordinates: $n = \mbox{const} > 1$. In this case $n$ is eliminated and we get the same trajectory and deflection angle (\ref{alpha-vacuum}) as in vacuum. The radius of photon circular orbits also will not change in comparison with vacuum.\\

\textit{Homogeneous dispersive medium in presence of gravity}.

This case is already discussed in the beginning of this Section. While there is no deflection in a homogeneous dispersive medium in the absence of gravity (Fig.\ref{fig:table}g), it occurs if gravity is added, and this deflection is chromatic (Fig.\ref{fig:table}h).\\

\textit{Non-homogeneous cold plasma.}

The propagation of light rays in the presence of gravity and cold plasma is well studied in the literature mentioned, so here we will focus only on certain properties that distinguish plasma from other media.

We consider cold, non-magnetized plasma with electron plasma frequency
\begin{equation}
\omega_p^2(r) = \frac{4 \pi e^2}{m_e} N(r) \, ,    
\end{equation}
where $e$ and $m_e$ are the electron charge and mass correspondingly, and $N(r)$ is the electron number density. 

The refractive index of such a plasma has the form (\ref{refr-index-plasma}). The Hamiltonian of such medium is reduced to the following form \cite{Perlick-2000, BK-Tsupko-2009, BK-Tsupko-2010, Tsupko-BK-2013, Perlick-Tsupko-BK-2015}:
\begin{equation}\label{Ham-plasma-1}
H(x^i, p_i) = \frac{1}{2} \left\{  g^{ik} p_i p_k + \omega_p^2(r) \right\}    \, .
\end{equation}
We find that for plasma the photon frequency $\omega(r)$ is eliminated from the Hamiltonian. In particular, this leads to the fact that the deflection angles do not depend on the velocity of the medium \cite{Perlick-2000}, because it can be included into the Hamiltonian by means of the formula (\ref{synge-freq-general}) only. This conclusion agrees, for example, with the consideration of light deflection by cold plasma stream in absence of gravity, see Ko and Chuang \cite{Ko-Chuang-1978}. They found that the normally incident ray passes through the moving cold
plasma without any bending of the ray direction.

For other media, the motion of the medium affects the propagation of light (we remind that in this article we restrict ourselves to the case of static media). Note also that the absence of the photon frequency $\omega(r)$ inside the Hamiltonian (\ref{Ham-plasma-1}) does not mean that the deflection angle become achromatic: the Hamiltonian contains the constant $p_0 = - \omega_0$.\\

\textit{Homogeneous cold plasma.}

An important special case is a homogeneous plasma. The number density, and hence the plasma frequency of such a plasma, are constants, $\omega_p = \mbox{const}$, and the refraction index is:
\begin{equation} \label{index-homogen}
n^2 = 1 - \frac{\omega_p^2}{\omega^2(r)} \, .
\end{equation}
The Hamiltonian takes the form:
\begin{equation}
H(x^i, p_i) = \frac{1}{2} \left\{  g^{ik} p_i p_k + \omega_p^2 \right\}    \, ,
\end{equation}
where $\omega_p$ is constant. This Hamiltonian can be compared with Hamiltonian describing the motion of massive particle in vacuum:
\begin{equation}
H(x^i, p_i) = \frac{1}{2} \left\{  g^{ik} p_i p_k + \mu^2 \right\}    \, ,
\end{equation}
where $\mu$ is the mass of test particle. As shown by Kulsrud and Loeb \cite{Kulsrud-Loeb-1992}, in a homogeneous plasma, photons move along exactly the same trajectories as massive particles in a vacuum, with an effective mass equal to the plasma frequency, with an energy equal to the photon energy, and with speed equal to the group velocity of light ray. This conclusion is valid only in homogeneous plasma, i.e. in absence of refraction.

As already discussed, the deflection of light in homogeneous plasma differs from vacuum case, because plasma is an example of dispersive medium. As shown in \cite{BK-Tsupko-2009, BK-Tsupko-2010}, the deflection angle by Schwarzschild black hole surrounded by homogeneous plasma can be written in approximation of weak deflection $(\hat{\alpha} \ll 1)$ as
\begin{equation} \label{grav-spectr}
\hat{\alpha} = \frac{2m}{R} \left( 1 + \frac{1}{1 - \omega_p^2/\omega_0^2} \right)  \, .
\end{equation}
It is worth to emphasize that the formula (\ref{grav-spectr}) contains the ratio $\omega_p^2/\omega_0^2$ whereas the refractive index of plasma (\ref{refr-index-plasma}) used in derivation of this formula contains $\omega_p^2/\omega^2(r)$.

If we additionally assume that $\omega_p^2/\omega_0^2 \ll 1$, then we find a correction to the vacuum gravitational deflection due to a homogeneous plasma:
\begin{equation} \label{plasma-angle-expanded}
\hat{\alpha} = \frac{4m}{R} \left( 1 + \frac{\omega_p^2}{2 \omega_0^2} \right) \, .
\end{equation}
In the next section, we will calculate a similar correction for an arbitrary dispersive medium.

\section{Application I: Correction to Einstein vacuum gravitational deflection due to presence of homogeneous dispersive medium}
\label{sec:correction-disp}

As we already discussed in the beginning of previous Section, in presence of gravitational field, the ray trajectories in the dispersive medium will differ from the  vacuum ones, even if the medium is homogeneous. Let us now calculate the correction related to homogeneous dispersive medium for the vacuum Einstein deflection.

We start with our general expression for deflection angle (\ref{alpha}) and apply it to the Schwarzschild spacetime,
\begin{equation}
A(r) = 1 - \frac{2m}{r} \, , \; B(r) = \left( 1 - \frac{2m}{r}  \right)^{-1} \, , \; D(r) = r^2 \, ,
\end{equation}
and refractive index $n(\omega)$. We deal with spatially homogeneous medium $n(\omega)$, so the refractive index does not depend explicitly on space coordinate $r$. In presence of gravity, we have the gravitational redshift (\ref{synge-freq-static}) for the photon frequency, so the refractive index reads as $n(\omega(r))$.

Let us expand the refractive index as
\begin{equation} \label{refr-index-exp}
n(\omega(r)) \simeq n_0 + n_1 \left( \omega(r) - \omega_0 \right) \, ,
\end{equation}
where we denote the values
\begin{equation}
n_0 = n(\omega_0) \, , \quad n_1 = \left. \frac{\partial n}{\partial \omega} \right|_{\omega=\omega_0} \, ,
\end{equation}
which do not depend on $r$. In absence of gravity, the second term in (\ref{refr-index-exp}) is absent, because $\omega=\omega_0$.

We further use the approximation that
\begin{equation} \label{expansion-cond}
\frac{n_1 \left( \omega(r) - \omega_0 \right)}{n_0} \ll 1 \, .
\end{equation}
in eq.(\ref{refr-index-exp}), and expand the deflection angle (\ref{alpha}). After that, the expressions are still big, and we consider the approximation of weak deflection of the light ray: $R \gg m$. Expanding with small variables $m/r \ll 1$ and $m/R \ll 1$, we can then perform integration analytically.

Finally, we obtain in approximation $\hat{\alpha} \ll 1$ that
\begin{equation}
\hat{\alpha} = \frac{4m}{R} \left( 1 + \frac{n_1 \omega_0}{2 n_0} \right) \, .
\end{equation}
We remind that the correction is present despite the fact that we deals with a homogeneous medium, that is, in the absence of refraction. Depending on the sign of the derivative $n_1$, the presence of a homogeneous medium can both increase and decrease the deflection angle.
For example, in a homogeneous plasma (\ref{index-homogen}), the deflection angle increases\cite{BK-Tsupko-2009, BK-Tsupko-2010}.

In particular case of homogeneous plasma (\ref{index-homogen}) we can approximately change the condition (\ref{expansion-cond}) by condition $\omega_p \ll \omega_0$. With this approximation in $n_0$ and $n_1$, we recover (\ref{plasma-angle-expanded}).

\section{Application II: Angular radius of black hole shadow}
\label{sec:shadow}

\begin{figure}
\begin{center}
\includegraphics[width=0.48\textwidth]{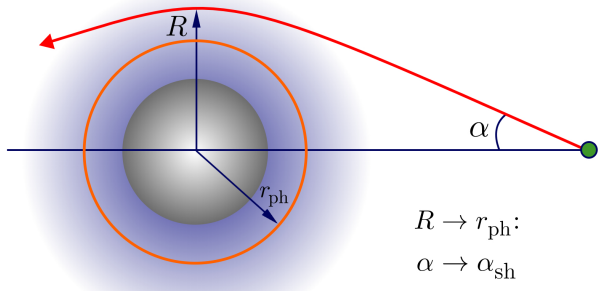}
\end{center}
\caption{Observer emits light ray into the past at angle $\alpha$ with respect to radial direction, and this ray passes by the black hole at the distance of closest approach $R$. If we substitute $R$ by the photon sphere radius $r_\textrm{ph}$, then the angle $\alpha$ becomes the angular radius $\alpha_\textrm{sh}$ of the black hole shadow.}
\label{fig:shadow}
\end{figure}

As another important example, we will calculate the angular size of the shadow of a black hole immersed in medium with a given refractive index. Black hole shadow 
\cite{Synge-1966, Zeld-Novikov-1965, Bardeen-1973, Chandra-1983, Dymnikova-1986, Falcke-2000, Takahashi-2004, Zakharov-Paolis-2005-New-Astronomy, Johannsen-2010, Johannsen-2013, Gren-Perlick-2014, Gren-Perlick-2015, Cunha-2015, Shipley-Dolan-2016, Tsupko-2017} has now become the active research topic due to the recent observational discovery \cite{EHT-1}, see, e.g., papers 
\cite{Eiroa-2018, Tsukamoto-2018, Cunha-Herdeiro-2018, Yunes-2018, Mars-2018, Perlick-Tsupko-BK-2018, BK-Tsupko-2018, Gralla-2019, Wei-2019-Rapid, Konoplya-2019, Alexeyev-2019, Psaltis-2019-review, Tsupko-Fan-BK-2020, Eubanks-2020, Dokuchaev-2020, Neves-2020, Farah-2020, Li-Guo-2020, Chang-Zhu-2020, Tsupko-BK-2020-IJMPD, Kumar-Ghosh-2020}. Using the same approach as in our previous paper \cite{Perlick-Tsupko-BK-2015}, we write down the angle $\alpha$ between direction of emission of light ray and radial direction (Fig.\ref{fig:shadow}):
\begin{equation} \label{shadow-01}
\cot \alpha = \frac{\sqrt{g_{rr}}}{\sqrt{g_{\varphi \varphi}}} \, \left. \frac{dr}{d\varphi} \right|_{r=r_\textrm{O}} = \frac{\sqrt{B(r)}}{\sqrt{D(r)}} \, \left. \frac{dr}{d\varphi} \right|_{r=r_\textrm{O}} \, .
\end{equation}
Here $r_\textrm{O}$ is the radial coordinate of observer.
Using (\ref{eq-orbit-R}) in (\ref{shadow-01}), we obtain:
\begin{equation}
\cot^2 \alpha = \frac{h^2(r_\textrm{O})}{h^2(R)} - 1 \, .
\end{equation}
This leads to
\begin{equation}
\sin^2 \alpha = \frac{h^2(R)}{h^2(r_\textrm{O})}  \, .
\end{equation}
The size of the shadow is determined by the rays that asymptotically spiral towards the photon sphere. Angular radius of the shadow $\alpha_\textrm{sh}$ is obtained if we put $R \to r_\textrm{ph}$:
\begin{equation} \label{shadow-final}
\sin^2 \alpha_\textrm{sh} = \frac{h^2(r_\textrm{ph})}{h^2(r_\textrm{O})}  \, .
\end{equation}
Here the radius $r_\textrm{ph}$ should be found from eq.(\ref{eq-circ}).

The final formula (\ref{shadow-final}) agrees with the formula derived for plasma case in \cite{Perlick-Tsupko-BK-2015}, with only difference in the definition of function $h(r)$, see (\ref{h-def}).
As with the deflection angle, we can see that the size of the shadow does not change when the refractive index is multiplied by a constant.

\section{Concluding remarks}
\label{sec:consclusions}

(i) We calculate the deflection angle of a light ray moving in an general spherically symmetric metric in the presence of a transparent static dispersive medium with a spherically symmetric refractive index, see eq.(\ref{alpha}) with (\ref{h-def}). The equation for the radius of the circular light orbits is also derived, see eq.(\ref{eq-circ}). The properties of these results and various special cases are discussed, see Sec.\ref{sec:discussion}.

(ii) We find that multiplying the refractive index by a constant does not affect the trajectories: the deflection angle and the radius of circular orbits don't change. In particular, this means that the trajectories of light rays in spatially homogeneous and non-dispersive medium are the same as in vacuum (see Fig.\ref{fig:table}e,f), although light velocity is smaller.

(iii) At the same time, the presence of dispersion leads to a difference of trajectories from the vacuum case, even in a spatially homogeneous medium. This is due to the fact that in the presence of a gravitational field, the frequency of the photon changes depending on the position in space according to the gravitational redshift (\ref{freq-relation}), and, accordingly, an effective inhomogeneity appears. As an application of our results, we calculate the correction to the vacuum gravitational deflection angle for the case when a massive object is surrounded by a homogeneous but dispersive medium, see Sec.\ref{sec:correction-disp}.

(iv) As another application, we present the calculation of the shadow of black hole surrounded by medium with an arbitrary refractive index, see Sec.\ref{sec:shadow}. This can be used for analytical calculation of shadow size beyond widely used cold plasma case, e.g., for different warm, hot or collisional plasma models.

\section*{Acknowledgements}
The author is grateful to Volker Perlick for many fruitful and pleasant discussions throughout work on this article.
Part of this paper was written during the research stay in the Hanse-Wissenschaftskolleg, Delmenhorst. OYuT is thankful to Kerstin Schill, Wolfgang Stenzel, Michael Kastner and Claus L\"{a}mmerzahl for great hospitality and help. The author brings special thanks to G.S. Bisnovatyi-Kogan for motivation and permanent help with all scientific initiatives.

\bibliographystyle{ieeetr}

\end{document}